# Molecular dynamics simulation of the capillary leveling of viscoelastic polymer films

I. Tanis[1], H. Meyer[2*], T. Salez[1,3], E. Raphaël[1], A.C. Maggs[1] and J. Baschnagel[2]

[1]Laboratoire de Physico-Chimie Théorique, UMR CNRS Gulliver 7083,
ESPCI Paris, PSL Research University, 75005 Paris, France

[2]Université de Strasbourg, CNRS, Institut Charles Sadron, UPR 22, 67000 Strasbourg, France

[3]Global Station for Soft Matter, Global Institution for Collaborative Research and Education,
Hokkaido University, Sapporo, Hokkaido 060-0808, Japan

*Author to whom correspondence should be addressed. Electronic mail:
hendrik.meyer@ics-cnrs.unistra.fr

**Abstract**. Surface tension-driven flow techniques have recently emerged as an efficient means of shedding light into the rheology of thin polymer films. Motivated by experimental and theoretical approaches in films bearing a varying surface topography, we present results on the capillary relaxation of a square pattern at the free surface of a viscoelastic polymer film, using molecular dynamics simulations of a coarse-grained polymer model. Height profiles are monitored as a function of time after heating the system above its glass-transition temperature and their time dependence is fitted to the theory of capillary leveling. Results show that the viscosity is not constant, but time dependent. In addition to providing a complementary insight about the local inner mechanisms, our simulations of the capillary-leveling process therefore probe the viscoelasticity of the polymer and not only its viscosity, in contrast to previous experimental approaches.

**I. Introduction**

Thin polymer films have emerged as a class of glass-forming materials with a potential for a wide number of nanoscale applications. Given the appealing properties such as (in general) low thermal conductivity and high dielectric constant, these systems are utilized in a variety of technological fields such as micro-electronics,[1,2] coatings in optical fibers[3] and nanolithography.[4] To this end, their static and dynamic properties and deviations from bulk behaviour due to spatial restriction and interfacial effects have been the subject of extensive research and debate over the last years.[5–8] Particular attention has been focused on the rheological behaviour of thin polymer films as well as on the dynamical heterogeneities triggered by the presence of interfaces.[9–12] A striking feature observed in films with a free interface is the enhanced mobility in a region near the surface, as suggested by recent reports.[9,13,14] Such behaviour has also been detected in small-molecule glasses.[15–20] An efficient and versatile method to probe the aforementioned attributes is the capillary-driven relaxation of an imposed topography on the film surface. In this context, a recent experimental approach addressed the viscocapillary leveling of polystyrene stepped films.[21] After heating well above the glass-transition temperature $T_g$, the excess surface area was allowed to relax under the effect of the Laplace pressure. By deriving the flow equations based on the lubrication approximation,[22] a measurement of the viscosity of the film was accomplished.[23,24] Furthermore, below $T_g$ the analysis suggests the existence of a thin mobile layer near the surface, whereas at elevated temperature the whole film flows.[25] The experimentally determined height profiles were in excellent agreement with the solution of the capillary-driven thin film equation which was found to converge in time to a self-similar profile.[23,24,26]

Molecular dynamics (MD) simulation has been a valuable tool for the elucidation of the mechanisms that determine the behaviour of polymers in confinement.[6,9,27–34] Given the wide spread of length and time scales that govern the structure and dynamics of such systems, different approaches and certain simplifications have been developed in order to meet the requirements of the system under study. The dramatic increase of the relaxation time upon approaching the glassy state is common to all polymers regardless of the underlying chemistry. This fact allows for the utilization of a simplified model with generic features of the polymer chain. Several studies employing such models have addressed interfacial effects on the glass transition and the alteration of the latter depending on film thickness, confinement-induced changes in polymer conformations and polymer-substrate interactions.[35–45] Smooth or rough walls with varying monomer-wall interaction strength have been used as substrates. A complex dynamical behaviour upon approaching the glass transition has been observed whereas shifts from the bulk $T_g$ were found in agreement with experimental measurements.[9,30] The spreading of droplets on viscous and glassy polymer films has been examined as well.[46] However, to the authors' knowledge, there is no simulation work considering polymer films with a stepped surface topography near the glass transition. As capillary leveling has been proven an efficient tool to probe surface mobility, the purpose of this contribution is to examine the evolution of thin supported films with a square surface pattern, using molecular dynamics simulations; thus complementing the existing experimental and theoretical approaches relying on a global knowledge of the surface profile with a newly accessible inner information at the molecular level. We monitor the evolution of the film and the leveling process at temperatures above $T_g$. Simulation results are fitted to an analytical solution of the Stokes equation assuming a small amplitude of the topographic perturbation. The latter assumption allows for a linearisation of the problem and application of Fourier analysis.[4,24] In section II the simulation model and the preparation of the polymer films are described whereas in section III we provide the analytical viscocapillary solution for a periodic square profile. Section IV describes our analysis and the results obtained from the simulation and analytical approaches. The final section summarizes our findings. An investigation of the polymer-substrate boundary condition as well as the local inner dynamics as a function of the distance from the wall, is presented in the Appendix. This examination considers either smooth, weakly-attractive substrates described implicitly by a wall potential or substrates modeled explicitly through a hexagonal lattice.

## II. Simulation protocol

A generic bead-spring (BS) model for a polymer melt which has been successfully utilized in previous studies of thin polymer films, is used for the molecular dynamics simulations in this work.[29,47] The model assumes linear, flexible chains for which only the connectivity and non-bonded monomer-monomer interactions for all but the nearest neighbours along the chain are taken into account. The connectivity between adjacent monomers of a chain separated by a distance $r$ is ensured by a harmonic-spring potential with equilibrium distance $l_b = 0.967\sigma$ and spring constant $k_b = 1111\ \varepsilon/\sigma^2$:

$$U_b = \frac{k_b}{2}(r - l_b)^2, \tag{1}$$

where $k_b$ has been chosen large enough in order that chains cannot cut through each other, thus allowing for the formation of entanglements. Non-bonded monomer-monomer interactions are described by a truncated and shifted Lennard-Jones (LJ) potential with cut-off radius, $r_c = 2.3\sigma \simeq 2r_{min}$ where $r_{min}$ is the distance corresponding to the minimum of the LJ potential:

$$U_{LJ}(r) = 4\varepsilon\left[\left(\frac{\sigma}{r}\right)^{12} - \left(\frac{\sigma}{r}\right)^6\right] - U_{LJ}(r_c),\ r \leq r_c, \tag{2}$$

where $\sigma$ stands for the monomer diameter and $\varepsilon$ for the depth of the potential well. This BS model is very similar to other models recently used in simulation studies of polymers in confinement.[27,31,39]

All simulation results are given in LJ units. The LJ parameters of the monomers as well as the monomer mass are set to unity. Length is measured in $\sigma$, energy in $\varepsilon$, temperature in units of $\varepsilon/k_B$, time in units of $\tau_{LJ} = \sqrt{m\sigma^2/\varepsilon}$, viscosity in units of $\tau_{LJ}\varepsilon/\sigma^3$ and surface tension in $\varepsilon/\sigma^2$. The model contains two intrinsic length scales, corresponding to the minima of the bond and LJ potentials. The existence of these distinct lengths in conjunction with the chain flexibility would allow us to assume that the system remains amorphous at low temperatures, as shown in several works.[31,43,44] Nevertheless, a recent study reports that bead-spring models widely used in systems in confinement, are subjected to heterogeneous nucleation when exposed to explicit crystalline substrates.[48] We should note here that no sign of crystallization was detected in our study. The polymer sample consists of a total number $n = 8064$ of chains, each bearing a number $N = 16$ of monomers. This chain length is below the estimated entanglement length, found to be $N_c \simeq 32$ [49,50] or $N_c \simeq 64$ based on primitive-path analysis.[51]

We examine systems of supported films with one polymer-substrate interface and one free (polymer-vacuum) interface. Two types of substrates were considered and described in detail in the Appendix. A smooth, weakly attractive structureless wall could be introduced to model the polymer-substrate interaction. However, as shown in the Appendix, this representation imparts a slip boundary condition together with some layering extending over several monomer diameters. In contrast, the incorporation of explicit substrate particles with perturbed coordinates with respect to a periodic lattice, was found to inhibit polymer slippage and restrict density modulation to a few monomer diameters. The latter case therefore constitutes an adequate representation of the established boundary condition for polystyrene on a silicon substrate in the experimental approaches.[21,23–26] Consequently, all films described below are supported on explicit substrates.

All simulations were conducted via the LAMMPS[52] code in the canonical (NVT) ensemble. As conservation of momentum is important for hydrodynamics, temperature was maintained through the use of the Dissipative Particle Dynamics (DPD) thermostat (pair friction)[53] which controls temperature via the balance of a random force that injects energy into the system and a friction force dissipating energy. For that thermostat, a coupling constant $\zeta = 1$ and a cutoff function $w(r) = 1 - \frac{r}{r_c}$, were used, with $r$ being the distance between a pair of monomers and the cutoff length $r_c$ being the same as that of the LJ potential.[53] The equations of motion were integrated by the Velocity Verlet algorithm with a time step of 0.005. Periodic boundary conditions were applied in the $xz$ directions, parallel to the substrate. In a previous work, the glass-transition temperature of supported and free-standing polymer films was estimated by monitoring the film thickness upon cooling.[47,54] The authors reported a minor reduction of $T_g$ ($T_g \approx 0.38$) for supported films of chain length $N = 10$ with respect to corresponding samples in the bulk ($T_g \approx 0.39$). Furthermore, for the purpose of the present study, additional runs performed on free-standing films of thicknesses $D = 5, 10, 20$ showed that for $D \geq 10$, mechanical stability was maintained for temperatures up to $T = 1$ and no "evaporation" of chains was observed.

In order to compare the simulated height profiles with the hydrodynamic theory, to be described later, estimations of the viscosity and surface tension are required. The former was initially calculated in the bulk state from the late-time value of the Green-Kubo formula: [55]

$$\eta = \lim_{t \to \infty} \eta(t) \qquad (3a)$$

$$\eta(t) = \int_0^t G(t')dt' \qquad (3b)$$

$$G(t) = \frac{V}{k_B T} \langle G_{ij}(t) G_{ij}(0) \rangle, \qquad (3c)$$

where $V$ stands for the volume of the system, $G_{ij}$ refers to the shear component of the stress tensor, the angle bracket denotes the ensemble average and $t$ is the time. Data analysis showed that $G(t)$ becomes noisy for $t \geq 1000$ as shown in fig.1. To remove the noise we utilized the Rouse model which, to a reasonable approximation, is valid for the examined chain length. On these grounds, $G(t)$ can be estimated by:[56]

$$G(t) = k_B T \rho \sqrt{1/(2\pi^2 Wt)} \, exp\left(-\frac{\pi^3}{4N^2} Wt\right), \qquad (4)$$

where $\rho$ denotes the monomer density, $N$ the chain length and $W$ the monomer relaxation rate, which was independently estimated to be equal to 0.000248 at $T = 0.44$ and 0.002309 at $T = 0.50$. As it can be seen in fig.1, the Rouse model provides an interpolation through the noise of the MD data and, in addition, allows for extrapolation to longer times. Using eqs.3 and integrating the merged (MD + Rouse) data one obtains the time-dependent bulk viscosity $\eta(t)$. In section IV we will perform a comparison of this bulk viscosity with fit results from the hydrodynamic model.

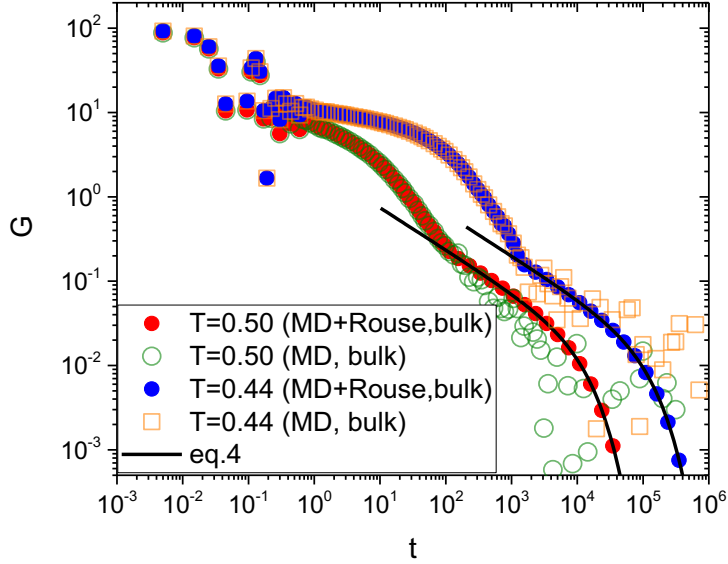

Figure 1: Shear-stress bulk autocorrelation function versus time $t$. The MD data is shown by open symbols. The solid lines indicate the Rouse model (eq.4) with $\rho = 1$. The filled circles show the merged data (MD + Rouse). Note that the oscillations at early times ($t < 0.1$) have a small contribution to the integral of $G(t)$.[57]

The surface tension was estimated via the local (per monomer) stress tensor. To this end, supported flat films of thickness $D \approx 20$ were prepared. Each average stress component can be extracted by counting the number of monomers in a layer at distance $y$ from the substrate, calculating the pressure for each monomer in the layer and afterwards averaging over all monomers in the layer.[58] Integration of the difference between the normal ($P_\perp$) and tangential ($P_\parallel$) components yields the surface tension:[47]

$$\gamma = \int_{D/2}^{\infty} [P_\perp(y) - P_\parallel(y)] dy . \qquad (5)$$

Our calculations yield $\gamma(T = 0.44) = 1.58$, $\gamma(T = 0.50) = 1.44$ and $P_\perp \approx 0$ for all $y$ as required by mechanical stability.

At the first stage of the patterned-film preparation, we introduced structureless walls at $y = 0$ and $y = B_y$ via eq.10 to a well relaxed flat film of thickness $D \approx 15$ with $n = 1152$ chains (box dimensions: $B_x = 52.4832$ and $B_y = B_z = 26.2416$) at a very low temperature of $T = 0.1$. This initial configuration was heated to $T = 0.50$. To do so, temperature was continuously increased at a rate of $2 \times 10^{-4}$. After completion of the heating ramp, we performed an isothermal run at $T = 0.50$. The length of this run was sufficiently long to achieve a relaxation of the root-mean-square value of the chain end-to-end distance. After quenching, this equilibrated configuration (without the two surrounding structureless walls) was used as base unit for the construction of the patterned film at $T = 0.1$. Two layers of substrate particles were inserted slightly above a structureless wall. The film base unit was replicated in the $x$ and $y$ directions in order to create the desired film profile atop the new substrate. This procedure finally yielded a periodic square pattern with a vertical aspect ratio of 1:3. The resulting topography is on average invariant in the $z$ direction. Similarly to the procedure described earlier, the patterned film was heated up to $T = 0.44$ and $0.50$. The configuration obtained after the heating ramp was used as the starting point for the isothermal runs and is illustrated in fig.2. For a number of the prepared systems, additional relaxation steps were performed for the bottom polymer layer (highlighted in blue in fig.2) to ensure the adhesion of the layer to the substrate and allow for some mixture of the chains. This pre-equilibration stage was found to have no significant effect on the leveling process.

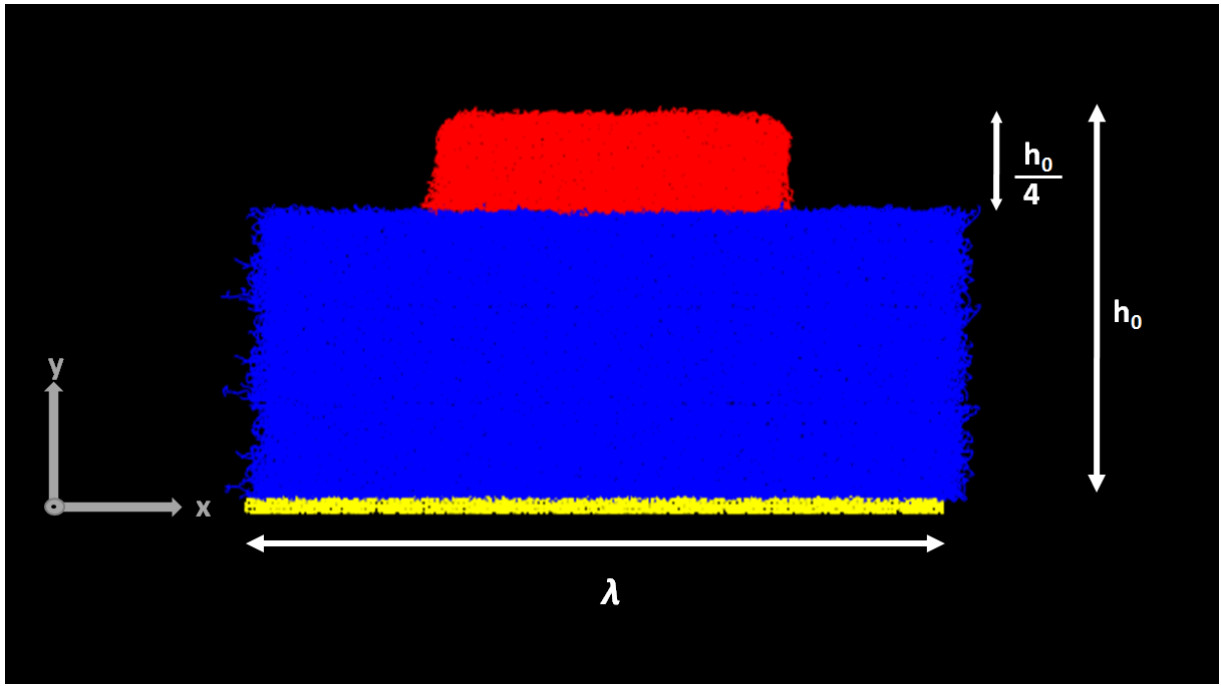

Figure 2: Configuration of the polymer-film model bearing a total number of $n = 8064$ chains with $N = 16$ monomers each, after the heating ramp to $T = 0.44$. Periodic boundary conditions are applied in the horizontal $x$, $z$ directions. $\lambda = 2B_x$ stands for the horizontal wavelength, i.e. the $x$-dimension of the simulation box and $h_0$ is the maximum vertical height (colour code : red = top layer, blue = bottom layer, yellow = substrate).

**III. Theory**

In order to characterize the profile evolution obtained from the MD simulations, we invoke the following hydrodynamic model. We consider a thin viscous film of viscosity $\eta$ and surface tension $\gamma$ with the ambient atmosphere. At $t = 0$, the initial film profile is a periodic symmetric square pattern identical to the one depicted in fig.2. At finite time $t$, the height profile is noted $h(x,t)$. Due to the chosen aspect ratio and invoking conservation of volume, the long-time fully-leveled flat state is characterized by a uniform height $h_\infty = \lim_{t\to+\infty} h(t) = 7h_0/8$.

By invoking low-Reynolds-number hydrodynamics with a no-slip boundary condition at the substrate and a no-shear boundary condition at the free interface and assuming that the surface perturbation is small, it is possible to show[4,59] that each Fourier mode of the profile with nonzero wavevector $k$ decays exponentially with a rate determined by $k$, $h_\infty$, $\gamma$, $\eta$ and $\lambda$. Introducing the dimensionless quantities $X = 2x/\lambda$, $K = kh_\infty$, $A = 2h_\infty/\lambda$ and $\delta_0 = 2(h_0 - h_\infty)/h_\infty = 2/7$, together with the dimensionless time:

$$T^* = \frac{t\gamma}{\eta h_\infty}, \tag{6}$$

one gets the dimensionless profile $H(X, T^*) = h(x,t) / h_\infty$ through:

$$H(X,T^*) = 1 + \frac{2\delta_0}{\pi} \sum_{n=0}^{\infty} \frac{\sin[(2n+1)\pi X]}{2n+1} e^{-\Omega([2n+1]\pi A)T^*}, \tag{7}$$

where we invoked the auxiliary function:

$$\Omega(K) = \frac{K}{2} \frac{e^{2K} - e^{-2K} - 4K}{e^{2K} + 2 + e^{-2K} + 4K^2}. \tag{8}$$

To verify that our simulation method satisfies the low-Reynolds number assumption of the hydrodynamic model, we evaluated the Reynolds number at the two target temperatures using the formula $\text{Re} = \rho h_0 u_c/\eta$, where $\rho$ is the monomer density and $u_c = \gamma/\eta$ the capillary velocity. Substitution of the numerical values of $\eta$, $\gamma$ and $h_0$ ($\eta(T = 0.44) \approx 5085$, $\eta(T = 0.50) \approx 560$; the calculations of $h_0$ are described in section IV) yields a Reynolds number on the order of $10^{-6}$ at $T = 0.44$ and $10^{-4}$ at $T = 0.50$ respectively, thus confirming the non-inertial nature of the simulated flow. In addition, the stress-tensor calculations described in section II yield a null value for the shear stress at the free surface, in agreement with the boundary condition invoked in the theory.

## IV. Analysis

As stated in section II, the configuration obtained after the heating ramp performed on the patterned film was used as the starting point for isothermal runs. The duration of each run was long enough to ensure substantial surface leveling. Typical temporal evolutions of the profile at both temperatures are presented in fig.3.

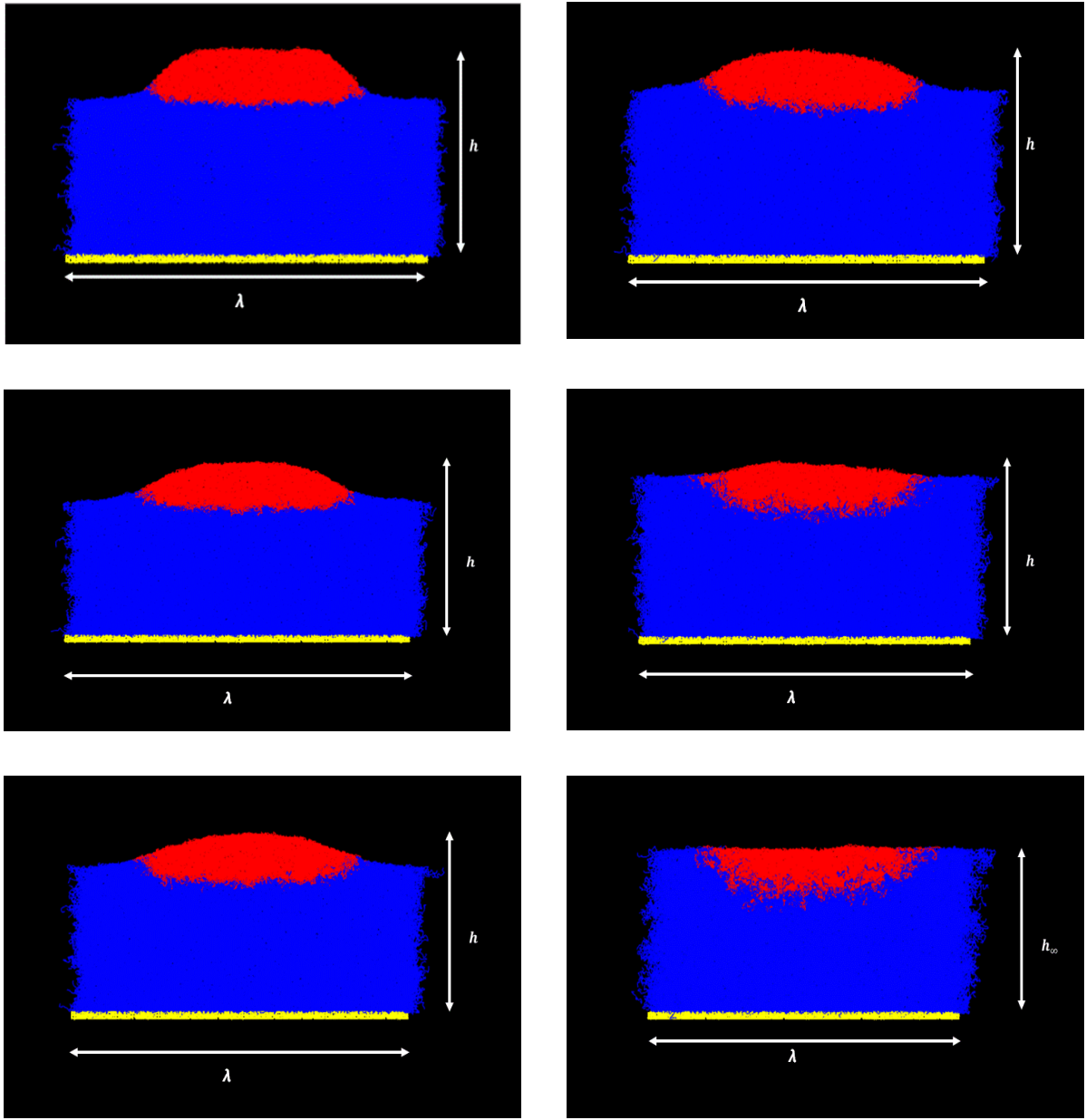

Figure 3: Time evolution of the square pattern (see fig.2) at $T = 0.44$ (left panels) and $T = 0.50$ (right panels). From top to bottom: $t/10^3 = 5, 20, 45$. Note that at the bottom right panel full leveling has been accomplished ($h = h_\infty$).

A clear qualitative difference in the film evolution is observed at the two different temperatures. This is anticipated as at $T = 0.44$ the film is close to the glass transition and chain motion remains slow whereas at $T = 0.50$ the system is essentially a less-viscous liquid. As can be seen from the right panels, complete leveling is reached at $T = 0.50$ in a time scale on the order of $50 \times 10^3$. On the other hand, at $T = 0.44$ complete leveling is seen (not shown) at about $330 \times 10^3$. The total simulation time reached approximately 750 CPU hours.

For a proper comparison with the theory, the *x*-origin of the MD profiles was defined through $h(x, t) / h_\infty = 1$. More precisely, the *x*-origin was determined on the first measured profile and then kept constant for all other times. With these corrections, $h_0$ and $h_\infty$ were found to be equal to 52.50 and 45.95, respectively, at *T* = 0.44, whereas the corresponding values at *T* = 0.50 were 53.31 and 46.64. Having determined $h_0$ and $h_\infty$, the only unknown in eqs.6 and 7 is *T\**. The latter was determined by fitting eq.7 to the simulation data. The height profiles obtained from the simulation data and from the fits to eq.7 are shown in fig.4 whereas table 1 lists the fit parameters for *T\** together with the effective viscosity $\eta_{film}(t)$ determined from eq.6.

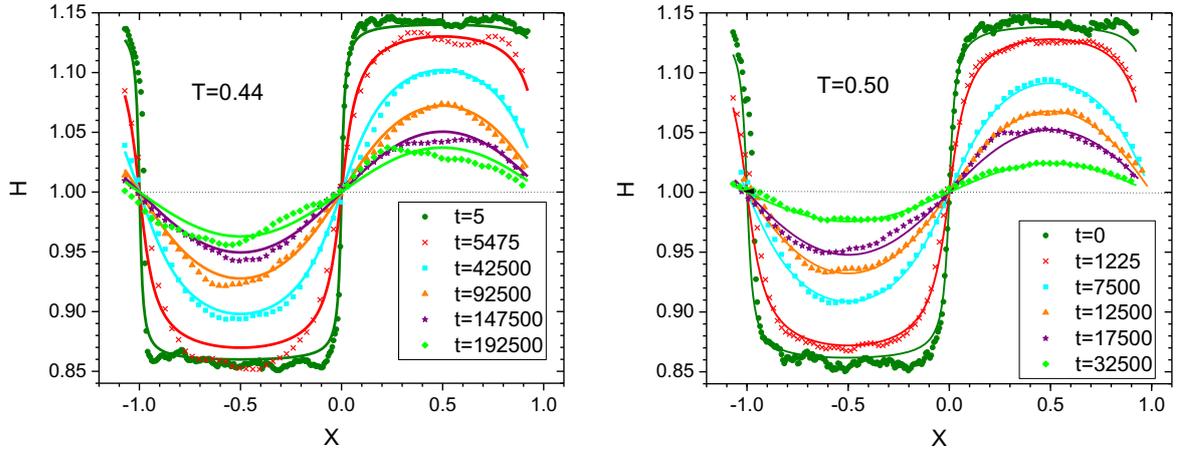

Figure 4: MD height profiles (symbols) at different times as indicated, for two temperatures: *T* = 0.44 (left) and *T* = 0.50 (right). The solid lines are fits to eq.7 with the dimensionless times *T\** as free parameters (see table 1). *X* = 0 corresponds to the position of the left side of the square pattern.

The excellent agreement between theory and MD data demonstrates the robustness of the simulation method. The visual inspection of the profiles also reveals an absence of "bumps" and "dips" on the respective thick and thin sides of the patterned film in contrast to experimental findings in stepped polystyrene films.[21] Moreover, an examination of the profiles at early times has not shown any clear indication of self-similar behavior, in contrast to the aforementioned case.[23] This is due to the fact that these previous studies were performed in the domain of validity of the lubrication approximation (for which the typical horizontal length scale is much larger than the vertical one) and that their initial patterns were not periodic. Invoking the lubrication approximation in our case would require $A = 2h_\infty/\lambda \ll 1$ whereas we instead have: $2h_\infty/\lambda \approx 1$. Analogous findings are reported in a combined experimental and theoretical study targeted for nanoimprint lithography.[4] On these grounds, eq.7 is more general and valid for our thick system.

Table 1: Fit parameters for the fits shown in fig. 4. Also listed for each temperature is the effective viscosity $\eta_{film}(t)$ deduced from $T^*$ and eq.6.

| T = 0.44 | | | T = 0.50 | | |
|---|---|---|---|---|---|
| t | $T^*$ | $\eta_{film}(t)$ | t | $T^*$ | $\eta_{film}(t)$ |
| 5 | 0.0282 | 6.0828 | 0 | 0.0492 | - |
| 5425 | 0.1348 | 1396.2 | 1225 | 0.1587 | 238.23 |
| 42500 | 0.4241 | 3445.1 | 7500 | 0.5293 | 437.43 |
| 92500 | 0.7645 | 4159.4 | 12500 | 0.8088 | 477.13 |
| 147500 | 1.0904 | 4650.3 | 17500 | 1.0388 | 520.09 |
| 192500 | 1.3610 | 4862.4 | 32500 | 1.6928 | 592.70 |

Focusing on table 1, we see that $\eta_{film}(t)$ determined from capillary leveling is found to depend on time. Therefore, our simulations appear to probe the viscoelasticity of the polymer by measuring how $\eta(t)$ converges to the steady-state viscosity. To test this idea, it is interesting to compare the effective viscosity $\eta_{film}(t)$ with the data obtained from the integration of $G(t)$ described in section II. Figure 5 displays the results from the two approaches.

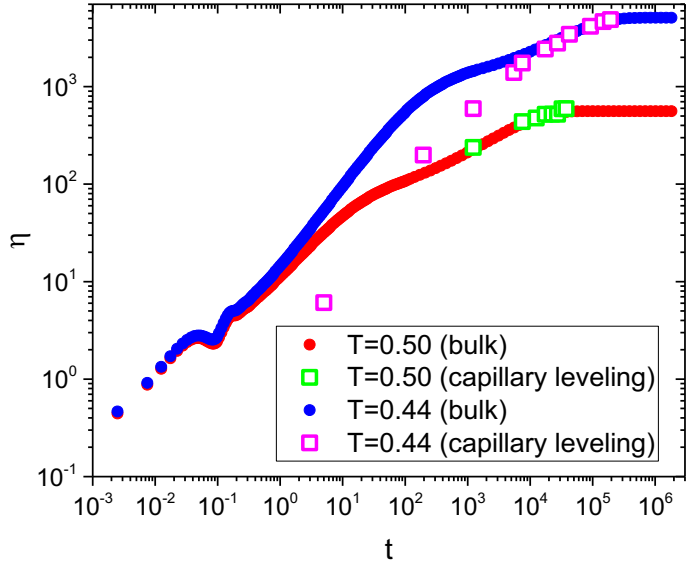

Figure 5: Bulk viscosity $\eta$ of the polymer (filled circles) as a function of time, extracted via eqs.3 by integrating the merged data for $G(t)$ obtained from the MD simulations and the Rouse model approximation. The open squares represent the capillary-levelling results for $\eta_{film}(t)$ listed in table 1.

Examining fig.5 we discern that the effective viscosity determined from capillary leveling, lies in good agreement with the bulk one after an initial transient regime, i.e. for $t \geq 1000$ at $T =$ 0.50 and $t \geq 10000$ at $T = 0.44$. This early regime is presumably affected by remnants of the sample preparation. It is also noteworthy that, unlike previous experimental approaches where the associated time scales range from tens of minutes to hours, our MD simulation probes a time window of the order of hundreds of nanoseconds and, consequently, the viscoelastic behaviour of the polymer film.

**IV. Conclusions**

Molecular dynamics simulations have been conducted in order to examine the capillary relaxation of a thin polymer film, the free surface of which bears a periodic square pattern. A generic bead-spring model was utilized to model the polymer chains. Concurrently, a hydrodynamic model, based on the Stokes equations, allowed to obtain an analytical prediction for the profile evolution in the limit of a small surface perturbation. The isothermal evolution of the height profiles was recorded after heating above the glass-transition temperature and was fitted to the analytical prediction. Excellent agreement is obtained under the assumption of a time-dependent viscosity. This is consistent with the simulation time scales which resolve the evolution before the longest chain relaxation time. However, we have to note that a detailed theory including this time dependence needs to be worked out. Upon approaching the glass transition, we expect that a height dependence of mobility and viscoelastic relaxation needs to be taken into account as well (the lower temperature in the current study was still 10% above the critical temperature predicted by the mode coupling theory).

As a conclusion, this work opens the way to the study of viscoelastic effects in confined complex fluids and polymer glasses via molecular dynamics simulation of capillary leveling.

**Acknowledgements**
The authors gratefully acknowledge financial support from ANR WAFPI and ANR FSCF, as well as the Global Station for Soft Matter, a project of Global Institution for Collaborative Research and Education at Hokkaido University. They thank Olivier Benzerara for computational support in doing the analysis shown in fig.4 as well as Kari Dalnoki-Veress, James Forrest, Joshua McGraw, and Oliver Bäumchen for stimulating discussions and experimental insights.

**Appendix. Hydrodynamic boundary condition: control and characterization**

As expected in a supported polymer film, the established hydrodynamic boundary condition is governed by the interaction of the polymer molecules with the substrate, together with the structure of the latter. Simulation tests were performed on a flat film of thickness $D \approx 20$. At an initial stage, we implicitly introduced a wall at $y = 0$. The monomer-wall interaction was modeled by a nontruncated 9-3 LJ-potential:

$$U_\text{w}(y) = \varepsilon_\text{w}\left[\left(\frac{\sigma}{y}\right)^9 - \left(\frac{\sigma}{y}\right)^3\right], \tag{9}$$

where $y$ stands for the distance from the wall and $\varepsilon_\text{w}$ for the potential depth. This formula can be obtained by integrating the LJ interactions between the wall atoms and a fixed test monomer.[60] To examine the effect of explicit substrate particles on the boundary condition, an additional wall model was generated. Two layers of substrate particles were inserted above the previous structureless wall. These particles were organized as a crystalline lattice but with positions randomly displaced by up to 25% of a lattice spacing. The size of the particles was chosen to be the same as the monomer size and the monomer-substrate interaction strength was chosen to be the same as that between the monomers. This choice of LJ parameters in conjunction to a monomer density $\rho \sim 1$, has been shown to enhance the ability of the polymer chains to adapt to the wall structure, thus leading to a no-slip boundary condition.[6,61,62] The substrate particles were fixed to their modified positions by stiff springs of constant $k_\text{s} = 200$ $\varepsilon/\sigma^2$. Both systems were generated at $T = 0.1$ and heated to $T = 0.50$ at a rate equal to the one stated in section II. Isothermal runs of 800000 steps were then launched at the target temperature. Figure 6 displays the monomer density profiles of the two examined systems at $T = 0.50$.

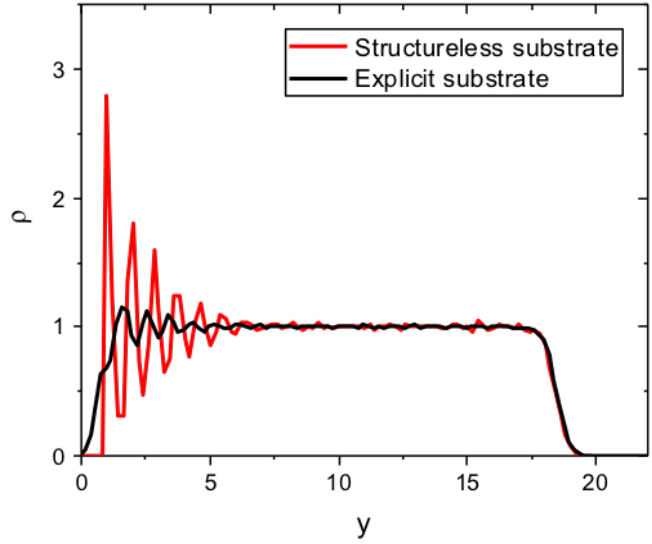

Figure 6: Monomer density profiles for a film supported on a wall represented either implicitly (eq.9) or by explicit substrate particles. The profile corresponding to the latter case has been shifted in the *y*-axis in order that the positions of the free surfaces of both systems coincide.

Figure 7 presents the layer-resolved incoherent intermediate scattering function calculated at $q = 6.9$, the position of the maximum of the static structure factor $S(q)$:

$$\varphi_q^{s\|}(t,y) = \left\langle \frac{1}{n_t} \sum_i \prod_{t'=0}^{t} \delta[y - y_i(t')] \, e^{-i\vec{q}\left[\vec{r}_i^{\|}(t) - \vec{r}_i^{\|}(0)\right]} \right\rangle. \qquad (10)$$

Here $\vec{r}_i^{\|}(t)$ is the position, parallel to the wall, of monomer *i* at time *t* and $\varphi_q^{s\|}$ is averaged over all monomers. This definition takes into account only the $n_t$ monomers which are at all times $t' < t$ within the slab centered at *y* and bearing a width of $\Delta y = 2$. We consider the dynamics in the directions *x* and *z*.

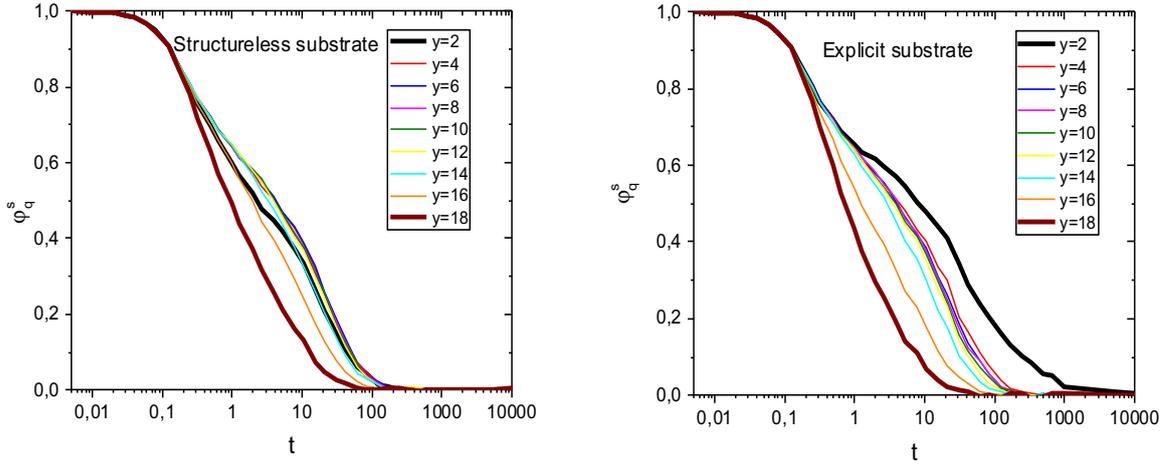

Figure 7: Layer-resolved incoherent scattering function at $q = 6.9$ (corresponding to the maximum of $S(q)$) and $T = 0.50$, with $y$ being the distance from the wall represented either implicitly (left) or by explicit substrate particles (right).

As it can be seen in fig.6, the implementation of explicit substrate particles results in a reduction of the region where density modulation is observed to few monomer diameters, together with a notable reduction of the amplitude of that modulation. Moreover, fig.7 reveals that the relaxation time increases as we move from a structureless wall to the centre of the film and decreases again as we approach the free surface.[54] In contrast, with the explicit substrate, dynamics gets slower upon approaching the wall, thus indicating a weak-slip boundary condition. This dynamical behaviour is in agreement with previous simulation studies in confined liquids.[6,27,63]

We further proceeded to the examination of the height profiles of films with a square surface pattern supported on both substrates. For reasons of computational efficiency, we considered a system with a 1:1 aspect ratio as illustrated in fig.8.

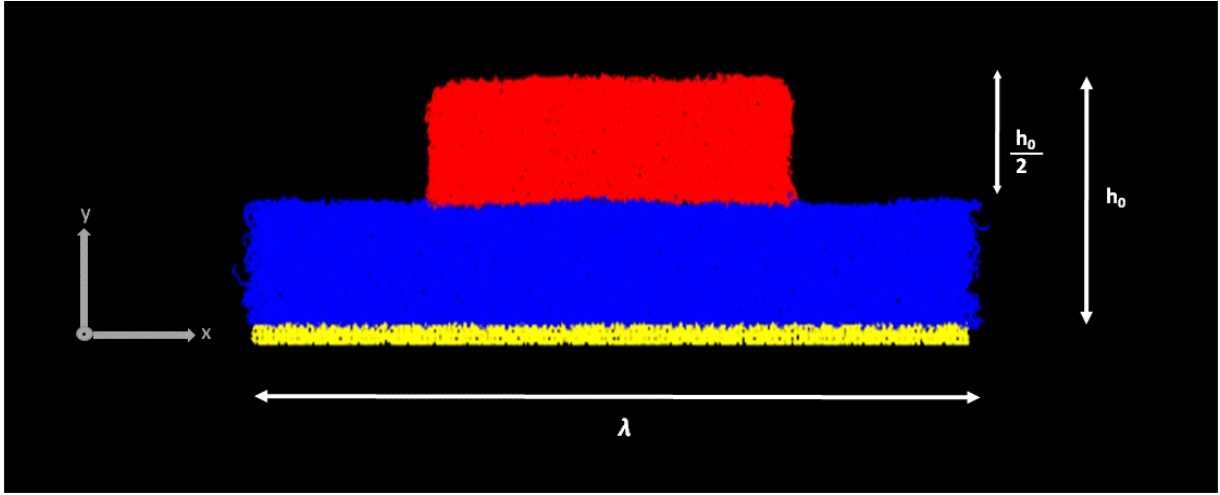

Figure 8: Polymer film with a square surface profile (periodic boundary conditions in the horizontal directions) placed here on an explicit substrate and bearing a total number of $n = 3456$ chains with $N = 16$ monomers each, after the heating ramp to $T = 0.44$. (colour code: red = top layer, blue = bottom layer, yellow = substrate).

The films were prepared and heated to $T = 0.44$ in a manner analogous to the procedure stated in the main text. The corresponding height profiles are shown in fig.9.

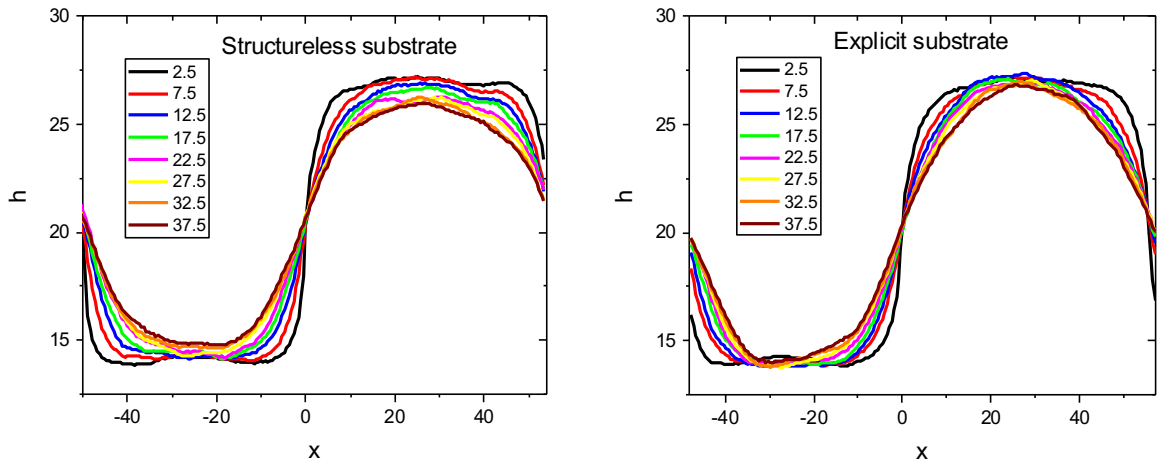

Figure 9: Height profiles at successive times ($t/10^3$ as indicated in legend) for structureless (left) and explicit (right) substrates. Similarly to fig.4, $x = 0$ corresponds to the position of the left side of the square pattern.

A different behaviour is observed in the height profiles, where the faster dynamics owing to the presence of the weakly attractive smooth wall (left) results in a quicker leveling.